\documentclass[preprint,amsmath,amssymb,11pt,english,aps,showpacs,showkeys]{revtex4}
\usepackage{graphicx}
\usepackage{graphics}
\usepackage{amsmath}
\usepackage{dcolumn}
\usepackage{amssymb}
\usepackage{bm}
\begin{document}
\title{An angular frequency dependence on the Aharonov-Casher geometric phase}
\author{P. M. T. Barboza}
\affiliation{Departamento de F\'isica, Universidade Federal da Para\'iba, Caixa Postal 5008, 58051-970, Jo\~ao Pessoa, PB, Brazil.}
\author{K. Bakke}
\email{kbakke@fisica.ufpb.br}
\affiliation{Departamento de F\'isica, Universidade Federal da Para\'iba, Caixa Postal 5008, 58051-970, Jo\~ao Pessoa, PB, Brazil.}

\begin{abstract}
A quantum effect characterized by a dependence of the angular frequency associated with the confinement of a neutral particle to a quantum ring on the quantum numbers of the system and the Aharonov-Casher geometric phase is discussed. Then, it is shown that persistent spin currents can arise in a two-dimensional quantum ring in the presence of a Coulomb-type potential. A particular contribution to the persistent spin currents arises from the dependence of the angular frequency on the geometric quantum phase. 
\end{abstract}

\keywords{Aharonov-Casher effect, magnetic dipole moment, geometric quantum phases, quantum ring, persistent spin current, biconfluent Heun function}
\pacs{03.65.Ge, 03.65.Vf}

\maketitle

\section{Introduction}

In recent decades, the Aharonov-Casher effect \cite{ac} has attracted a great deal of attention. This quantum effect arises in the quantum dynamics of a neutral particle with a permanent magnetic dipole moment and consists in a phase shift in the wave function of the neutral particle \cite{anan,anan2}. This phase shift arises from the interaction between the permanent magnetic dipole moment of the neutral particle and an electric field, and it is given by
\begin{eqnarray}
\phi_{\mathrm{AC}}=\oint\left(\vec{\mu}\times\vec{E}\right)\cdot\,d\vec{r}=\pm2\pi\mu\lambda,
\end{eqnarray} 
where $\mu$ is the magnetic dipole moment and $\lambda$ is a constant associated with a linear electric charge density. Discussions about the Aharonov-Casher effect have been extended to the nondispersivity and nonlocality \cite{ac7}, and also to the topological nature \cite{ac9}. The Aharonov-Casher effect \cite{ac} has also been investigated in recent years in the Lorentz-symmetry violation background \cite{belich,lbb}, in the noncommutative quantum mechanics \cite{fur2}, in noninertial reference frames \cite{bf4,extra2,extra3,extra4}, in holonomic quantum computation \cite{holo} and with respect to its dual effect, which is called in the literature as the He-McKellar-Wilkens effect \cite{hmw}. The gravitational analogue of the Aharonov-Casher effect was obtained in Ref. \cite{extra}.

Geometric quantum phases \cite{i1,i2} have also been studied in mesoscopic systems where there exists a dependence of the energy levels on the geometric phase \cite{dloss}. This dependence on the geometric phase gives rise to a quantum effect called as the Aharonov-Bohm effect for bound states \cite{ab,peshkin}. Moreover, this flux dependence of the energy levels yields the appearance of persistent currents in mesoscopic systems \cite{dloss,by,ring,fur4}. Based on a neutral particle system that possesses a permanent magnetic dipole moment, then, persistent spin currents have been studied in Refs. \cite{ring7,bf15}.

In this paper, persistent spin currents in a two-dimensional quantum ring are investigated. It is shown that bound states can be obtained for a neutral particle possessing a permanent magnetic dipole moment confined to a two-dimensional quantum ring in the presence of a Coulomb-type potential. By considering the interaction between the permanent magnetic dipole moment of the neutral particle and a radial electric field, then, three quantum effects can be observed in this system: one is the dependence of the energy levels on the Aharonov-Casher geometric phase; the second quantum effect is a dependence of the angular frequency associated with the confinement of a neutral particle to a quantum ring on the quantum numbers of the system and also on the geometric quantum phase; the third quantum effect is the arising of persistent spin currents in the two-dimensional quantum ring. A particular contribution to the persistent spin currents comes from the dependence of the angular frequency on the geometric quantum phase. It is worth mentioning that this last contribution has not been reported in the literature yet thus, as an example, we obtain both angular frequency and the persistent spin current associated with the ground state of the system.

The structure of this paper is as follows: in section II, we discuss the quantum dynamics of neutral particle that possesses a permanent magnetic dipole moment confined to a two-dimensional quantum ring in the presence of a Coulomb-type potential. We show that the energy levels depend on the Aharonov-Casher geometric phase, and also show a dependence of the angular frequency on the quantum numbers of the system and on the geometric quantum phase. Finally, we obtain both angular frequency and the persistent spin current associated with the ground state of the system; in section III, we present our conclusions.

\section{quantum dynamics of a spin-half neutral particle}

Let us start with the relativistic quantum dynamics of a neutral particle with permanent magnetic dipole moment that interacts with external magnetic and electric fields. This quantum dynamics can be described by introducing a nonminimal coupling into the Dirac equation as $i\gamma^{\mu}\partial_{\mu}\rightarrow i\gamma^{\mu}\partial_{\mu}+\frac{\mu}{2}\,F_{\mu\nu}\left(x\right)\,\Sigma^{\mu\nu}$, where $F_{\mu\nu}\left(x\right)$ is the electromagnetic tensor, with $F_{0i}=E_{i}$, $F_{ij}=-\epsilon_{ijk}\,B^{k}$ and $\mu$ is the permanent magnetic dipole moment of the neutral particle \cite{ac,anan,anan2}. In particular, in the Aharonov-Casher system \cite{ac}, the magnetic moment of the neutral particle interacts with an electric field produced by a linear distribution of electric charges, $\vec{E}=\frac{\lambda}{\rho}\,\hat{\rho}$, where $\rho=\sqrt{x^{2}+y^{2}}$, and $\lambda$ is a constant associated with the linear charge distribution along the $z$ axis. Based on this symmetry, from now on, we need to deal with the Dirac equation in cylindrical coordinates (curvilinear coordinates). The mathematical formulation used to write the Dirac equation in curvilinear coordinates is the same of spinors in curved spacetime \cite{weinberg,bd,schu}. Then, in cylindrical coordinates, the line element of the Minkowski spacetime is writing in the form: $ds^{2}=-dt^{2}+d\rho^{2}+\rho^{2}d\varphi^{2}+dz^{2}$ ($\hbar=c=1$). Therefore, by applying a coordinate transformation $\frac{\partial}{\partial x^{\mu}}=\frac{\partial \bar{x}^{\nu}}{\partial x^{\mu}}\,\frac{\partial}{\partial\bar{x}^{\nu}}$ and a unitary transformation on the wave function $\psi\left(x\right)=U\,\psi'\left(\bar{x}\right)$, the Dirac equation for a neutral particle with permanent magnetic dipole moment can be written in any orthogonal system in the following form \cite{schu,lbb}: $i\,\gamma^{\mu}\,D_{\mu}\,\Psi+\frac{i}{2}\,\sum_{k=1}^{3}\,\gamma^{k}\,\left[D_{k}\,\ln\left(\frac{h_{1}\,h_{2}\,h_{3}}{h_{k}}\right)\right]\Psi+\frac{\mu}{2}\,F_{\mu\nu}\left(x\right)\,\Sigma^{\mu\nu}\Psi=m\Psi$, where $D_{\mu}=\frac{1}{h_{\mu}}\,\partial_{\mu}$ is the derivative of the corresponding coordinate system and the parameter $h_{k}$ corresponds to the scale factors of this coordinate system. In our case (cylindrical coordinates), the scale factors are $h_{0}=1$, $h_{1}=1$, $h_{2}=\rho$ and $h_{3}=1$. Further, if the neutral particle is subject to a scalar confining potential $V$, then, the general form of the Dirac equation can be written as $i\,\gamma^{\mu}\,D_{\mu}\,\Psi+\frac{i}{2}\,\sum_{k=1}^{3}\,\gamma^{k}\,\left[D_{k}\,\ln\left(\frac{h_{1}\,h_{2}\,h_{3}}{h_{k}}\right)\right]\Psi+\frac{\mu}{2}\,F_{\mu\nu}\left(x\right)\,\Sigma^{\mu\nu}\Psi=\left[m+V\left(\rho\right)\right]\Psi$. However, our focus in this work is on a nonrelativistic neutral confined to a mesoscopic system, therefore, we need to take the nonrelativistic limit of the Dirac equation above. We can take the nonrelativistic limit of the Dirac equation by using, for instance, the Foldy-Wouthuysen approximation up to terms of order $m^{-1}$ \cite{greiner,fur2,bf15}. In this way, we obtain the Schr\"odinger-Pauli equation (by working with 2-spinors and the units $\hbar=c=1$) \cite{bf15}:
\begin{eqnarray}
i\frac{\partial\psi}{\partial t}=\frac{1}{2m}\left(\vec{p}+\vec{\Xi}\right)^{2}\psi-\frac{\mu^{2}E^{2}}{2m}\,\psi+\frac{\mu}{2m}\left(\vec{\nabla}\cdot\vec{E}\right)\psi+\mu\vec{\sigma}\cdot\vec{B}\,\psi+V\left(\rho\right)\,\psi.
\label{1}
\end{eqnarray}
The matrices $\vec{\sigma}$ are the Pauli matrices, where they obey the relation $\left(\sigma^{i}\,\sigma^{j}+\sigma^{j}\,\sigma^{i}\right)=2\,\eta^{ij}$, with $\eta^{ab}=\mathrm{diag}(+ + +)$. In addition, the components of the vector $\vec{\Xi}$ are given by $\Xi_{k}=\mu\left(\vec{\sigma}\times\vec{E}\right)_{k}+\frac{1}{2}\,\sigma^{3}\,e^{\varphi}_{\,\,\,k}\left(x\right)$ \cite{bf15}. The term $\frac{1}{2}\,\sigma^{3}\,e^{\varphi}_{\,\,\,k}\left(x\right)$ is the contribution that stems from the spinorial connection. By the way, in the Aharonov-Casher system \cite{ac}, the term $\mu\left(\vec{\sigma}\times\vec{E}\right)=\pm\frac{\phi_{\mathrm{AC}}}{2\pi\rho}\,\hat{\varphi}$ plays the role of an effective vector potential, and yields the appearance of a phase shift in the wave function of the neutral particle which is called as the Aharonov-Casher effect \cite{ac}. Recently, two interesting studies of neutral particle systems have been made in Refs. \cite{extra5,extra6}.

In what follows, let us investigate the behaviour of a neutral particle that interacts with a radial electric field confined to a two-dimensional quantum ring by assuming that this system is subject to a Coulomb-type potential. An interest in including a Coulomb-type potential comes from the studies of quark models \cite{quark,eug}, the Kratzer potential \cite{kratzer}, the Mie-type potential \cite{molecule}, position-dependent mass systems \cite{pdm2,pdm3,pdm5} and topological defects in solids \cite{bm,mil,kit,rei,vis}. Recently, the two-dimensional Coulomb potential has been investigated under the presence of the Aharonov-Bohm effect \cite{cab,cab2,cab3}. According to Ref. \cite{quark}, the introduction of a Coulomb-like potential in the Dirac equation as an additional term of the mass of the particle has a particular interest in studies of confinement of quarks, such as in the {\it bag} model, since it can generalize the rest mass of the particle to the model of confinement of quarks. On the other hand, the two-dimensional quantum ring is a mesoscopic system described by the Tan-Inkson model \cite{tan}, then, by considering the presence of a Coulomb-type potential, the scalar potential $V\left(\rho\right)$ in Eq. (\ref{1}) is written in the form:
\begin{eqnarray}
V\left(\rho\right)=\frac{\eta}{\rho}+\frac{a_{1}}{\rho^{2}}+a_{2}\,\rho^{2}+V_{0},
%\label{3.1}
\label{2}
\end{eqnarray}
where $V_{0}=\sqrt{a_{1}\,a_{2}}$ and the parameters $\eta$, $a_{1}$ and $a_{2}$ are constant parameters. Note that, by taking the parameter $\eta=0$, the expression (\ref{2}) represents the Tan-Inkson model for a two-dimensional quantum ring \cite{tan}. By observing that the motion of the neutral particle inside the two-dimensional quantum ring is in the region where $\rho\neq0$, then, the term $\vec{\nabla}\cdot\vec{E}$ is null by assuming that the wave function of the neutral particle is well-behaved at the origin. Besides, we consider the permanent magnetic dipole moment of the neutral particle to be parallel to the $z$ axis, hence, the Schr\"odinger-Pauli equation (\ref{1}) becomes
\begin{eqnarray}
i\frac{\partial\psi}{\partial t}&=&-\frac{1}{2m}\left[\frac{\partial^{2}\psi}{\partial\rho^{2}}+\frac{1}{\rho}\frac{\partial\psi}{\partial\rho}+\frac{1}{\rho^{2}}\frac{\partial^{2}\psi}{\partial\varphi^{2}}+\frac{\partial^{2}\psi}{\partial z^{2}}\right]+\frac{i}{2m}\frac{\sigma^{3}}{\rho^{2}}\frac{\partial\psi}{\partial\varphi}-\frac{1}{2m\rho^{2}}\frac{\phi_{\mathrm{AC}}}{2\pi}\psi\nonumber\\
[-2mm]\label{3}\\[-2mm]
&-&\frac{i}{m}\,\frac{\phi_{\mathrm{AC}}}{2\pi}\frac{\sigma^{3}}{\rho^{2}}\,\frac{\partial\psi}{\partial\varphi}+\frac{1}{2m\rho^{2}}\left(\frac{\phi_{\mathrm{AC}}}{2\pi}\right)^{2}\,\psi+\frac{1}{8m\rho^{2}}\psi+\frac{\eta}{\rho}\,\psi+\frac{a_{1}}{\rho^{2}}\,\psi+a_{2}\,\rho^{2}+V_{0}\,\psi.\nonumber
\end{eqnarray}

We can observe that $\psi$ is an eigenfunction of $\sigma^{3}$, whose eigenvalues are $s=\pm1$. Moreover, we have that the total angular momentum $\hat{J}_{z}=-i\partial_{\varphi}$ \cite{schu} and the operator $\hat{p}_{z}=-i\partial_{z}$ commute with the Hamiltonian operator given in right-hand side of Eq. (\ref{3}). Thereby, a particular solution to the Schr\"odinger-Pauli equation (\ref{3}) can be written in terms of the eigenfunctions of the operators $\hat{J}_{z}$ and $\hat{p}_{z}$, that is, $\psi_{s}=e^{-i\mathcal{E}t}\,e^{i\left(l+\frac{1}{2}\right)\varphi}\,e^{ikz}\,f_{s}\left(\rho\right)$, where $l=0,\pm1,\pm2,\ldots$ and $k$ is a constant. Thus, substituting $\psi_{s}$ into Eq. (\ref{3}), we obtain
\begin{eqnarray}
\left[\frac{d^{2}}{d\rho^{2}}+\frac{1}{\rho}\frac{d}{d\rho}-\frac{\tau^{2}}{\rho^{2}}-2m\,a_{2}\,\rho^{2}-\frac{2m\eta}{\rho}+\beta\right]f_{s}\left(\rho\right)=0,
\label{3.4}
\end{eqnarray}
where we have defined the parameters: $\tau^{2}=\gamma_{s}^{2}+2m\,a_{1}$, $\gamma_{s}=l+\frac{1}{2}\left(1-s\right)+s\frac{\phi_{\mathrm{AC}}}{2\pi}$ and $\beta=2m\left(\mathcal{E}-V_{0}-\frac{k^{2}}{2m}\right)$. Now, let us simplify our system by taking $k=0$ and perform a change of variables given by: $\xi=\left(2ma_{2}\right)^{1/4}\,\rho$; thus, we obtain 
\begin{eqnarray}
f_{s}''+\frac{1}{\xi}\,f_{s}'-\frac{\tau^{2}}{\xi^{2}}\,f_{s}-\frac{\delta}{\xi}\,f_{s}-\xi^{2}\,f_{s}+\frac{\beta}{\sqrt{2m\,a_{2}}}\,f_{s}=0,
\label{3.2}
\end{eqnarray}
where $\delta=\frac{2m\eta}{\left(2ma_{2}\right)^{1/4}}$. Observe that the asymptotic behaviour of the possible solutions to Eq. (\ref{3.2}) are determined for $\xi\rightarrow0$ and $\xi\rightarrow\infty$. Based on Refs. \cite{eug,mhv,vercin,heun}, we have that the behaviour of the possible solutions to Eq. (\ref{3.2}) at $\xi\rightarrow0$ and $\xi\rightarrow\infty$ allows us to write the function $f_{s}\left(\xi\right)$ in terms of an unknown function $H_{s}\left(\xi\right)$ as:
\begin{eqnarray}
f_{s}\left(\xi\right)=e^{-\frac{\xi^{2}}{2}}\,\xi^{\left|\tau\right|}\,H_{s}\left(\xi\right).
\label{3.3}
\end{eqnarray}
Substituting (\ref{3.3}) into (\ref{3.2}), we obtain
\begin{eqnarray}
H_{s}''+\left[\frac{2\left|\tau\right|+1}{\xi}-2\xi\right]H_{s}'+\left[g-\frac{\delta}{\xi}\right]H_{s}=0,
\label{3.4}
\end{eqnarray}
where $g=\frac{\beta}{\sqrt{2ma_{2}}}-2-2\left|\tau\right|$. The function $H_{s}\left(\xi\right)$ is a solution to the second order differential equation (\ref{3.4}) and it is known as the biconfluent Heun function \cite{heun,eug,bm}: $H_{s}\left(\xi\right)=H\left(2\left|\tau\right|,\,0,\,\frac{\beta}{\sqrt{2ma_{2}}},\,2\delta,\,\xi\right)$. We proceed with our discussion about bound states solutions by using the Frobenius method \cite{arf,f1,eug}. Therefore, the solution to Eq. (\ref{3.4}) can be written as a power series expansion around the origin: $H_{s}\left(\xi\right)=\sum_{j=0}^{\infty}\,d_{j}\,\xi^{j}$. Substituting this series into Eq. (\ref{3.4}), we obtain the following recurrence relation:
\begin{eqnarray}
d_{j+2}=\frac{\delta}{\left(j+2\right)\,\left(j+1+\theta\right)}\,d_{j+1}-\frac{\left(g-2j\right)}{\left(j+2\right)\,\left(j+1+\theta\right)}\,d_{j},
\label{3.12}
\end{eqnarray}
where $\theta=2\left|\tau\right|+1$. By starting with $d_{0}=1$ and using the relation (\ref{3.12}), we can calculate other coefficients of the power series expansion $H_{s}\left(\xi\right)=\sum_{j=0}^{\infty}\,d_{j}\,\xi^{j}$. For example, we have
\begin{eqnarray}
d_{1}&=&\frac{\delta}{\theta};\nonumber\\
[-2mm]\label{3.13}\\[-2mm]
d_{2}&=&\frac{\delta^{2}}{2\theta\left(1+\theta\right)}-\frac{g}{2\left(1+\theta\right)}.\nonumber
\end{eqnarray}

Observe that we can obtain bound state solutions by imposing that the power series expansion $H_{s}\left(\xi\right)=\sum_{j=0}^{\infty}\,d_{j}\,\xi^{j}$ becomes a polynomial of degree $n$. Through the recurrence relation (\ref{3.12}), this power series expansion becomes a polynomial of degree $n$ by imposing the conditions:
\begin{eqnarray}
g=2n\,\,\,\,\,\,\mathrm{and}\,\,\,\,\,\,d_{n+1}=0,
\label{3.13a}
\end{eqnarray}
where $n=1,2,3,\ldots$, and $g=\frac{\beta}{\sqrt{2ma_{2}}}-2\left|\tau\right|-2$. By analysing the condition $g=2n$, we obtain the energy levels for bound states:
\begin{eqnarray}
\mathcal{E}_{n,\,l,\,s}=\omega\left[n+\left|\tau\right|+1\right]+V_{0},
\label{3.14}
\end{eqnarray}
where $\omega=\sqrt{\frac{2a_{2}}{m}}$ is the corresponding angular frequency of the system. The energy levels (\ref{3.14}) corresponds to the spectrum of energy of a neutral particle with a permanent magnetic dipole moment confined to a two-dimensional quantum ring in the presence of a Coulomb-type potential. In contrast to recent studies of confinement of particles to a quantum ring \cite{tan,bf15,fur4}, the presence of the Coulomb-type potential modifies the spectrum of energy of the confinement of a particle to a two-dimensional quantum ring, where the ground state is determined by the quantum number $n=1$ instead of the quantum number $n=0$. Besides, the angular frequency of the Tan-Inkson model $\omega_{0}=\sqrt{\frac{8a_{2}}{m}}$ \cite{tan} is modified by the influence of the Coulomb-type potential, where the angular frequency of the system becomes $\omega=\sqrt{\frac{2a_{2}}{m}}$. Now, let us analyse the condition $d_{n+1}=0$ given in Eq. (\ref{3.13a}). Recently, the analysis of the condition $d_{n+1}=0$ made in Refs. \cite{bm,eug} has yielded a relation between the angular frequency and the quantum numbers of the system. For instance, in Ref. \cite{bm}, a relation involving a coupling constant of a Coulomb-like potential, the cyclotron frequency and the total angular momentum quantum number in semiconductors threaded by a dislocation density is obtained. In Ref. \cite{eug}, it is obtained a relation involving the mass of a relativistic particle, a scalar potential coupling constant and the total angular momentum quantum number. From the mathematical point of view, the relation in which involves the angular frequency and the quantum numbers $\left\{n,\,l,\,s\right\}$ results from the fact that the exact solutions to Eq. (\ref{3.4}) are achieved for some values of angular frequency.

Henceforth, let us discuss the relation between the angular frequency $\omega=\sqrt{\frac{2a_{2}}{m}}$ and the quantum numbers of the system. First of all, let us assume that the parameter $a_{2}$ of the Tan-Inkson model \cite{tan} can be adjusted in such a way that the condition $d_{n+1}=0$ is satisfied. The meaning of the assumption is that some specific values of the parameter $a_{2}$ are allowed in order to obtain bound state solutions. Moreover, the allowed values of $a_{2}$ depend on the quantum numbers $\left\{n,\,l,\,s\right\}$; thus, we label $a_{2}=a_{2}^{n,\,l,\,s}$ and the angular frequency is written as $\omega=\omega_{n,\,l,\,s}$. Hence, the two conditions established in Eq. (\ref{3.13a}) are satisfied and a polynomial expression to the function $H_{s}\left(\xi\right)$ is achieved. Let us exemplify this discussion by considering the ground state, which is determined by $n=1$. In this case, the condition $d_{n+1}=0$ yields $d_{2}=0$, then, by using the relation (\ref{3.13}), the corresponding angular frequency associated with the ground state is given by 
\begin{eqnarray}
\omega_{1,\,l,\,s}=\frac{2\,m\,\eta}{\left(2\left|\tau\right|+1\right)}.
\label{3.15}
\end{eqnarray}

In this way, the general expression for the energy levels (\ref{3.14}) must be written as:
\begin{eqnarray}
\mathcal{E}_{n,\,l,\,s}=\omega_{n,\,l,\,s}\,\left[n+\left|\tau\right|+1\right]+\sqrt{a_{1}\cdot a_{2}^{n,\,l\,s}}\,.
\label{3.16}
\end{eqnarray}

Hence, from Eqs. (\ref{3.15}) and (\ref{3.16}) we can see that the effects of the Coulomb-type potential on the spectrum of energy of the neutral particle confined to a two-dimensional quantum ring corresponds to a change of the energy levels, whose ground state is defined by the quantum number $n=1$ and the angular frequency depends on the quantum numbers $\left\{n,\,l,\,s\right\}$. This dependence of the angular frequency on the quantum numbers $\left\{n,\,l,\,s\right\}$ means that not all values of the angular frequency are allowed, but some specific values of the parameter $a_{2}$ which are defined in such a way that the conditions established in Eq. (\ref{3.13a}) are satisfied and a polynomial solution to the function $H_{s}\left(\xi\right)$ is achieved \cite{eug,f1,b50,bm}. Note that, even though there exists the presence of a Coulomb-like potential, this system remains a mesoscopic system and, therefore we can expect the same scale of energy of that discussed in Ref. \cite{tan}. 

Besides, due to the parameter $\tau^{2}=\left[l+\frac{1}{2}\left(1-s\right)+s\frac{\phi_{\mathrm{AC}}}{2\pi}\right]^{2}+2ma_{1}$, we can observe in Eqs. (\ref{3.15}) and (\ref{3.16}) the dependence of both energy levels and the angular frequency on the Aharonov-Casher geometric phase \cite{ac}. This dependence of the energy levels on the geometric quantum phase yields the arising of persistent spin currents in the quantum ring \cite{by,ring7,bf15}. In order that an expression of the persistent spin currents can be obtained, let us also consider a particular case given by the ground state. Thereby, by considering the ground state ($n=1$), the energy level is  
\begin{eqnarray}
\mathcal{E}_{1,\,l,\,s}=\frac{2\,m\,\eta}{\left(2\left|\tau\right|+1\right)}\,\left[\left|\tau\right|+2\right]+\frac{\sqrt{2m^{3}\,\eta^{2}\,a_{1}}}{\left(2\left|\tau\right|+1\right)},
\label{3.16a}
\end{eqnarray}
where we have substitute the angular frequency (\ref{3.15}) into Eq. (\ref{3.16}). Note that the energy level of the ground state (\ref{3.16a}) depends on the Aharonov-Casher geometric quantum phase with periodicity $\phi_{0}=\pm2\pi$, then, we have that $\mathcal{E}_{1,\,l,\,s}\left(\phi_{\mathrm{AC}}+\phi_{0}\right)=\mathcal{E}_{1,\,l+1,\,s}\left(\phi_{\mathrm{AC}}\right)$. By following Refs. \cite{by,fur4,tan,ring,ring7,fur5}, the expression for the total persistent spin currents that arises in the ring is given by $\mathcal{I}=\sum_{n,\,l}\mathcal{I}_{n,\,l}$, where $\mathcal{I}_{n,\,l}$ is defined with respect to current density operator  $\vec{J}$ as $\mathcal{I}_{n,\,l}=\left\langle J_{\varphi}\right\rangle=\left\langle \psi_{n\,l}\right|\hat{p}_{\varphi}+\Xi_{\varphi}\left|\psi_{n\,l}\right\rangle$. It is also shown that the expression for $\mathcal{I}_{n,\,l}$ can be simplified by writing it as $\mathcal{I}_{n,\,l}=-\frac{\partial\mathcal{E}_{n,\,l}}{\partial\phi}$ (where $\phi$ is the geometric quantum phase), which is called as the Byers-Yang relation \cite{by}. Therefore, since the energy level (\ref{3.16a}) depends on the Aharonov-Casher geometric phase $\phi_{\mathrm{AC}}$, the persistent spin current associated with the ground state of the system is
\begin{eqnarray}
\mathcal{I}_{1,\,l}=-\frac{\partial\mathcal{E}_{1,\,l,\,s}}{\partial\phi_{\mathrm{AC}}}=\frac{s}{\pi}\frac{\gamma_{s}}{\left|\tau\right|}\frac{1}{\left(2\left|\tau\right|+1\right)}\left[\frac{2m\eta\left(2+\left|\tau\right|\right)}{\left(2\left|\tau\right|+1\right)}-m\eta+\frac{\sqrt{2m^{3}\,\eta^{2}\,a_{1}}}{\left(2\left|\tau\right|+1\right)}\right].
\label{3.17}
\end{eqnarray}

Hence, we can observe that a quantum effect that corresponds to the arising of persistent spin currents in the two-dimensional quantum ring. In particular, the dependence of the angular frequency on the Aharonov-Casher quantum phase yields new contributions to the persistent spin current associated with the ground state in contrast to that obtained in Ref. \cite{bf15}. We must also observe that the expressions for both the angular frequency and the persistent spin current change for other values of the quantum number $n$, that is, for other energy levels.

\section{conclusions}

By investigating the effects of a Coulomb-type potential on the confinement of the Aharonov-Casher system \cite{ac} to a two-dimensional quantum ring \cite{tan}, we have seen that three quantum effects can be observed in this system: one is the dependence of the energy levels on the Aharonov-Casher geometric phase; the second quantum effect is a dependence of the angular frequency on the quantum numbers of the system and also the geometric quantum phase; the third quantum effect is the arising of persistent spin currents in the two-dimensional quantum ring. As a particular case, we have obtained the angular frequency of the ground state and calculated the persistent spin current associated with the ground state.

\acknowledgments

The authors would like to thank the Brazilian agencies CNPq and CAPES for financial support.

\end{document}